\documentclass[namedreferences,hyperref,optionalrh,solaromanenum]{spr-sola}
\usepackage{comment}
\usepackage{graphicx}                    
\usepackage{color}                       
\usepackage{siunitx}

\newcommand{\aap}{{\it Astron. Astrophys.}}

\newcommand{\apj}{{\it Astrophys. J.}}
\newcommand{\apjl}{{\it Astrophys. J. Lett.}}

\newcommand{\jgr}{{\it J. Geophys. Res.}}

\newcommand{\solphys}{{\it Sol. Phys.}}
 
\newcommand{\ssr}{{\it Space Sci. Rev.}} 
\newcommand{\araa}{{\it Annual Review of Astron and Astrophysis}}
\chardef\us=`\_

\usepackage{breakurl}        
\usepackage{rotating}
\usepackage[normalem]{ulem}

\newcommand\wsa{\texttt{WSA}}
\newcommand\euhforia{\texttt{EUHFORIA}}

\begin{document}

\begin{frontmatter}

\title{Evaluating Solar Wind Forecast Using Magnetic Maps
That Include Helioseismic Far-Side Information}

\author[addressref={UH,UGraz},corref,email={stephan.heinemann@hmail.at, stephan.heinemann@helsinki.fi}]{\inits{S.G.H.}\fnm{Stephan G.}~\lnm{Heinemann}\orcid{0000-0002-2655-2108}}

\author[addressref=MPS]{\inits{D.Y.}\fnm{Dan}~\lnm{Yang}\orcid{0000-0001-7570-1299}}

\author[addressref=NASA]{\inits{S.I.J.}\fnm{Shaela I.}~\lnm{Jones}\orcid{0000-0000-0000-0000}}

\author[addressref=UH]{\inits{J.P.}\fnm{Jens}~\lnm{Pomoell}\orcid{0000-0003-1175-7124}}

\author[addressref=UH]{\inits{E.A.}\fnm{Eleanna}~\lnm{Asvestari}\orcid{0000-0002-6998-7224}}

\author[addressref=AFRL]{\inits{C.J.H.}\fnm{Carl J.}~\lnm{Henney}\orcid{0000-0002-6038-6369}}%

\author[addressref=NASA]{\inits{C.N.A.}\fnm{Charles N.}~\lnm{Arge}\orcid{0000-0003-1662-3328}}

\author[addressref={MPS,UG,UAE}]{\inits{L.G.}\fnm{Laurent}~\lnm{Gizon}\orcid{0000-0001-7696-8665}}

\address[id=UH]{Department of Physics, University of Helsinki, P.O. Box 64, 00014, Helsinki, Finland}
\address[id=UGraz]{Institute of Physics, University of Graz, Universitätsplatz 5, 8010 Graz, Austria}
\address[id=MPS]{Max-Planck-Institut für Sonnensystemforschung, Justus-von-Liebig-Weg 3, 37077 G\"ottingen, Germany}
\address[id=NASA]{Heliophysics Science Division, NASA Goddard Space Flight Center, Code 671, Greenbelt, MD,
20771, USA}
\address[id=AFRL]{Air Force Research Laboratory, Space Vehicles Directorate, Kirtland AFB, NM 87117, USA}

\address[id=UG]{Institut f\"ur Astrophysik und Geophysik, Georg-August-Universit\"at G\"ottingen,  37077~G\"ottingen, Germany}
\address[id=UAE]{Center for Astrophysics and Space Science, NYUAD Institute, New York University Abu Dhabi, Abu Dhabi, UAE}

\runningauthor{Heinemann et al.}
\runningtitle{Evaluating Solar Wind Forecast Using FARM}

\begin{abstract}
To model the structure and dynamics of the heliosphere well enough for {high-}quality forecasting, it is essential to accurately estimate the global solar magnetic field used as inner boundary condition in {solar wind} models. However, our understanding of the photospheric magnetic field topology is inherently constrained by the limitation of systematically observing the Sun from only one vantage point, Earth. To address this challenge, we introduce global magnetic field maps that assimilate far-side active regions derived from helioseismology into solar wind modeling. Through a comparative analysis between the {combined surface flux transport and helioseismic Far-side Active Region Model} (FARM) {magnetic maps} and the base surface flux transport model without far-side active regions (SFTM), we assess the feasibility and efficacy of incorporating helioseismic far-side information in space weather forecasting. We are employing the Wang-Sheeley-Arge Solar Wind (\wsa) model for statistical evaluation and leveraging the EUropean Heliospheric FOrecasting Information Asset (\euhforia), a three-dimensional heliospheric MHD model, to analyze a case study. Using the \wsa\ model, we show that including far-side magnetic data improves solar wind forecasts for 2013–2014 by up to $50\%$ in correlation and $3\%$ in {root mean square error and mean absolute error,} especially near Earth and Solar TErrestrial RElations Observatory - Ahead (STEREO-A). Additionally, our 3D modeling shows significant {localized} differences in heliospheric structure that can be attributed to the presence or absence of active regions in the magnetic maps used as input boundaries. This highlights the importance of including far-side information to {more} accurately model and predict space weather effects caused by solar wind, solar transients, and geomagnetic disturbances.
\end{abstract}

\end{frontmatter}

\section{Introduction}
     \label{S-Introduction} 

The continuous stream of plasma emitted by the Sun, known as the solar wind, forms the foundation of heliospheric science, particularly in understanding solar-terrestrial interactions, i.e., space weather \citep{schwenn06,Temmer2021}. Accurate prediction of solar wind behavior is crucial for the operation and protection of satellites, power grids, communication systems, and crewed missions in low-Earth orbit \citep{2022Buzulukova}.\\

The solar wind, which originates from the rapid expansion of the solar corona \citep{1958parker, 2019cranmer}, is commonly understood to exhibit a bimodal nature. On the one hand, there is the so-called slow solar wind generated via processes mediated by reconnection \citep[although currently there exists no unanimous consensus regarding its origin;][]{Temmer2023}, such as interchange reconnection in the S-web model \citep[separatrix and quasi-separatrix web;][]{Antiochos2011} and closed field reconnection as observed in active region cusps, among other possible sources. On the other hand, the fast solar wind component is believed to originate in solar coronal holes \citep[][and references therein]{cranmer2002}.\\

As the solar wind flows outward from the Sun in a nearly radial direction, it stretches the coronal magnetic field. Due to the Sun's rotation, in the interplanetary space the magnetic field takes on an approximately Archimedean spiral shape, the so-called \textit{Parker} spiral \citep{1958parker}. Consequently, the interaction between slow and fast solar wind forms stream interaction regions (SIRs) that may evolve into recurrent co-rotating interaction regions (CIRs) \citep[{e.g.} see][]{jian06,2018richardson} that are the major cause of weak to moderate geomagnetic activity at Earth \citep[][]{1997farrugia,alves06,verbanac11}. \\

The geoeffectiveness of coronal mass ejections (CMEs) can be influenced by the solar wind structuring the heliosphere\citep{2024Heinemann}. When CMEs interact with the solar wind, particularly when they encounter high-speed solar streams (HSSs) or SIRs/CIRs, they may undergo significant modifications as they journey through the heliosphere. These alterations can cause the embedded flux rope to deform, twist, rotate or be deflected \citep[as observed in studies by][]{2004manchester, 2004riley, 2008Yurchyshyn,2013Isavnin,2019Heinemann_hsscme}. The flux rope may also erode due to reconnection processes \citep[][]{2006Dasso, 2012ruffenach,2015ruffenach, 2014lavraud, 2018wangY} or be deflected \citep[][]{2004wang,2014wangY, 2013kay,2019zhuang}. These interactions can also lead to increased turbulence within the sheath region \citep[]{2015lugaz,kilpua17}. All of these effects collectively alter the initial characteristics of CMEs observed near the Sun, and consequently predicting their arrival times and geoeffectiveness becomes a complex undertaking \citep[][]{2018richardson}. This highlights the critical role of the ambient solar wind structure in studying and forecasting space weather effects.\\

Solar wind prediction models rely on diverse data sources, and {an important factor in their accuracy is the} quality and comprehensiveness of the input data.
Different methods incorporate various datasets for forecasting purposes and many solar and heliospheric solar wind models, such as the Wang-Sheeley-Arge \citep[\wsa:][]{2000arge} model, Enlil \citep{Odstrvcil1999}, or the EUropean Heliospheric FOrecasting Information Asset \citep[\euhforia:][]{pomoell2018}, require an instantaneous estimate of the global solar (photospheric) magnetic field as a prerequisite for constructing the coronal magnetic field model. However, concurrent magnetic field observations of the entire Sun are not available on a regular basis. The common approach of using synoptic charts as an approximation of the instantaneous photospheric full-Sun magnetic field exhibits a drawback known as the ``{aging} effect''\citep{2021Heinemann}. As synoptic maps can only provide $360^{\circ}$ longitudinal information over the span of one solar rotation ($\approx27$ days), some parts are distinctly older than others. Although the inclusion of magnetic field observations from viewpoints away from the Earth-Sun line, for example those captured by the Photospheric and Magnetic Imager \citep[PHI:][]{2020solanki_so_phi} on board the Solar Orbiter \citep[SolO:][]{2020muller_solO} are highly beneficial, they are not consistently available.\\

An alternative approach to mitigate the {aging} effect involves the utilization of surface flux transport models, that attempt to capture the evolution of the large-scale photospheric magnetic field due to meridional flows and differential rotation, along with magnetic diffusion {(e.g. ADAPT\footnote{Air Force Data Assimilation Photospheric Flux Transport model}: \citealt{2010arge}; AFT\footnote{Advective Flux Transport model}: \citealt{AFT2014}; OFT\footnote{Open-source Flux Transport model}: \citealt{2025caplan_oft})}. However, these models typically do not account for the emergence of new magnetic flux on the Sun's {far-side}, and if so, only stochastically. Consequently, emerging flux and even entirely new active regions are only observed when the Sun rotates into Earth's field-of-view. This causes ``outdated'' or inaccurate magnetic maps that are used for modeling the solar wind. As a solution proposed by \cite{2013Arge} and subsequently implemented by \cite{Yang2024_FARM}, the incorporation of far-side information derived from helioseismology into magnetic charts was considered in this study. Using solar oscillations, we can infer spherical symmetric averages of the composition, density, and temperature at different depths inside the Sun. Far-side helioseismology utilizes acoustic waves that make multiple refractive skips between the surface and interior of the Sun, where regions of strong magnetic fields influence the effective travel time for a given acoustic wave. The differences in travel times around the solar near-surface interior make it possible to deduce magnetic information about the {far-side} of the Sun \citep[][]{Lindsey2000,Braun2001, 2007Gonzlez, Yang2023}{}{}. {In a recent study, \cite{2024Upton} utilized EUV observations as a proxy for far-side active regions and incorporated them into the AFT model, demonstrating that this approach improves the representation of the far-side magnetic field.}\\

In this study, we evaluate the value and feasibility of including far-side information in magnetic maps for space weather forecasting and heliospheric modeling.  We analyze \wsa\ solar wind predictions over a two-year period in the ecliptic plane and then further examine changes in the three-dimensional (3D) heliospheric solar wind structure in a case study, focusing on their implications for ambient solar wind forecasting and CME propagation. Through this analysis, we aim to assess the potential of helioseismology to contribute to the advancement of space weather modeling.

\section{Data and Methodology}
      \label{S-Statistics}   


\begin{figure*}[h]
   \centering
   \includegraphics[width=0.9\linewidth]{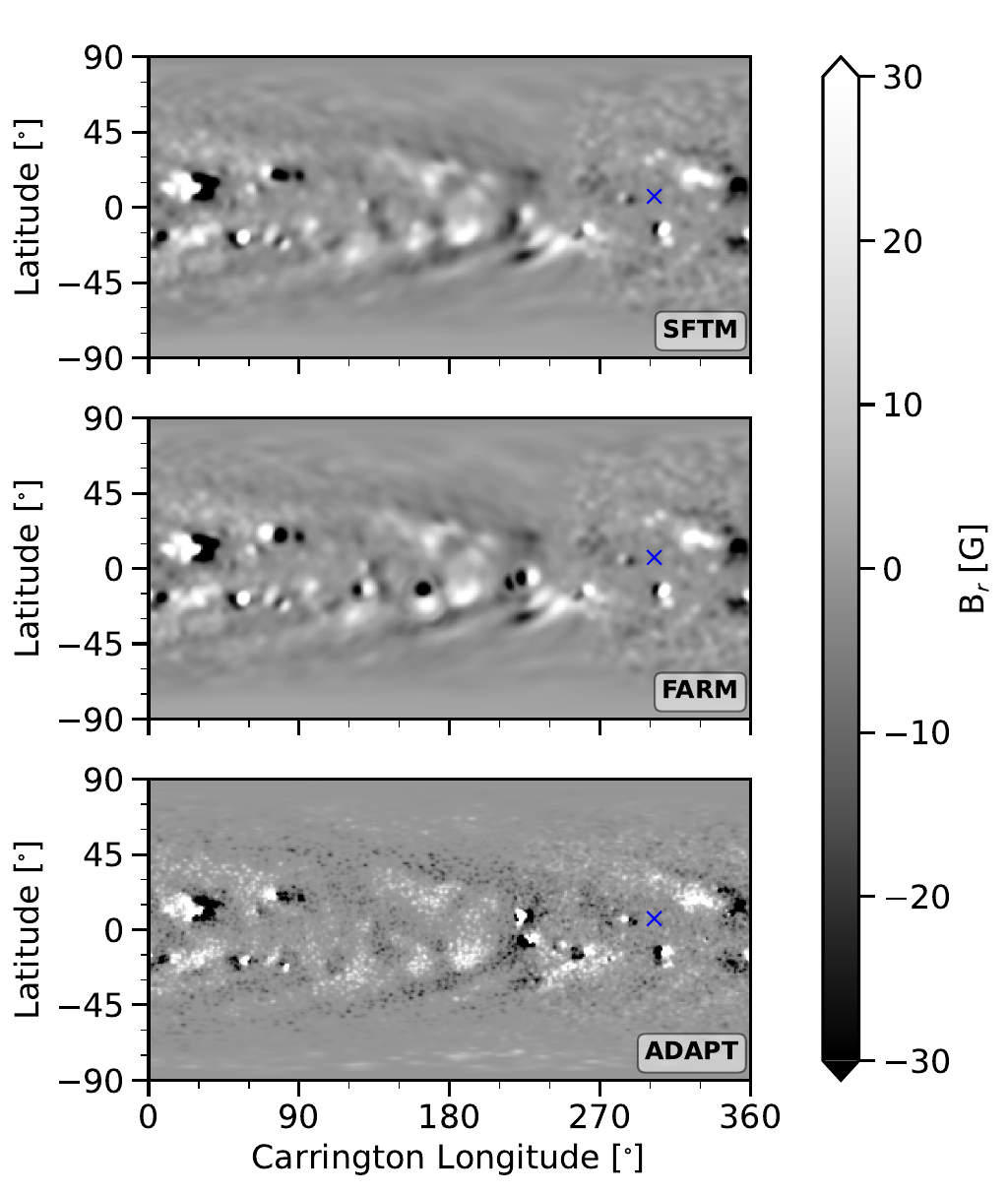}
   \caption{Sample magnetic maps from 2 October 2013 that show the SFTM (before inclusion of far-side active regions), FARM (with far-side active regions included), and ADAPT (reference global magnetic maps without far-side sources). The position of Earth’s longitude is marked by a blue 'x'.}
              \label{fig:magnetograms}%
\end{figure*}

To assess the impact of incorporating helioseismic far-side active regions into global photospheric magnetic maps, we analyze solar wind predictions using the semi-empirical \wsa\ solar wind model \citep[][also see Appendix~\ref{sec:wsa}]{wang90,2000arge} using maps from the {combined surface flux transport and helioseismic Far-side Active Region Model} \citep[FARM:][]{Yang2024_FARM}. We compare these maps to predictions based on maps without far-side data, generated by the baseline Surface Flux Transport Model (SFTM) on which FARM is based on. The dataset, produced by \citet{Yang2024_FARM}, spans the period from {2010} to 2024, during which data from the Solar Dynamics Observatory \citep[SDO;][]{2012pesnell_SDO} is available. The data is provided at a resolution of $1^{\circ}$ in both longitude and latitude, with a daily cadence. Further details on the magnetic maps can be found in \citet{Yang2024_FARM}.\\

In addition to the newly developed FARM maps, we also use magnetic maps generated by the well-established Air Force Data Assimilative Photospheric Flux Transport (ADAPT) model \citep{2010arge,Hickmann2015,Barnes2023}, which assimilates data from the Helioseismic and Magnetic Imager \citep[HMI:][]{2012schou_HMI} on board SDO, to serve as a reference. ADAPT provides 12 realizations for each time, based on variations in supergranular flows. We compute \wsa\ predictions for the full ensemble to estimate the associated uncertainty range. ADAPT maps are used at the same resolution and cadence as FARM and SFTM maps. Figure~\ref{fig:magnetograms} shows example maps from ADAPT, FARM, and SFTM for 2 October 2013.\\

\begin{figure*}[h]
   \centering
   \includegraphics[width=1\linewidth]{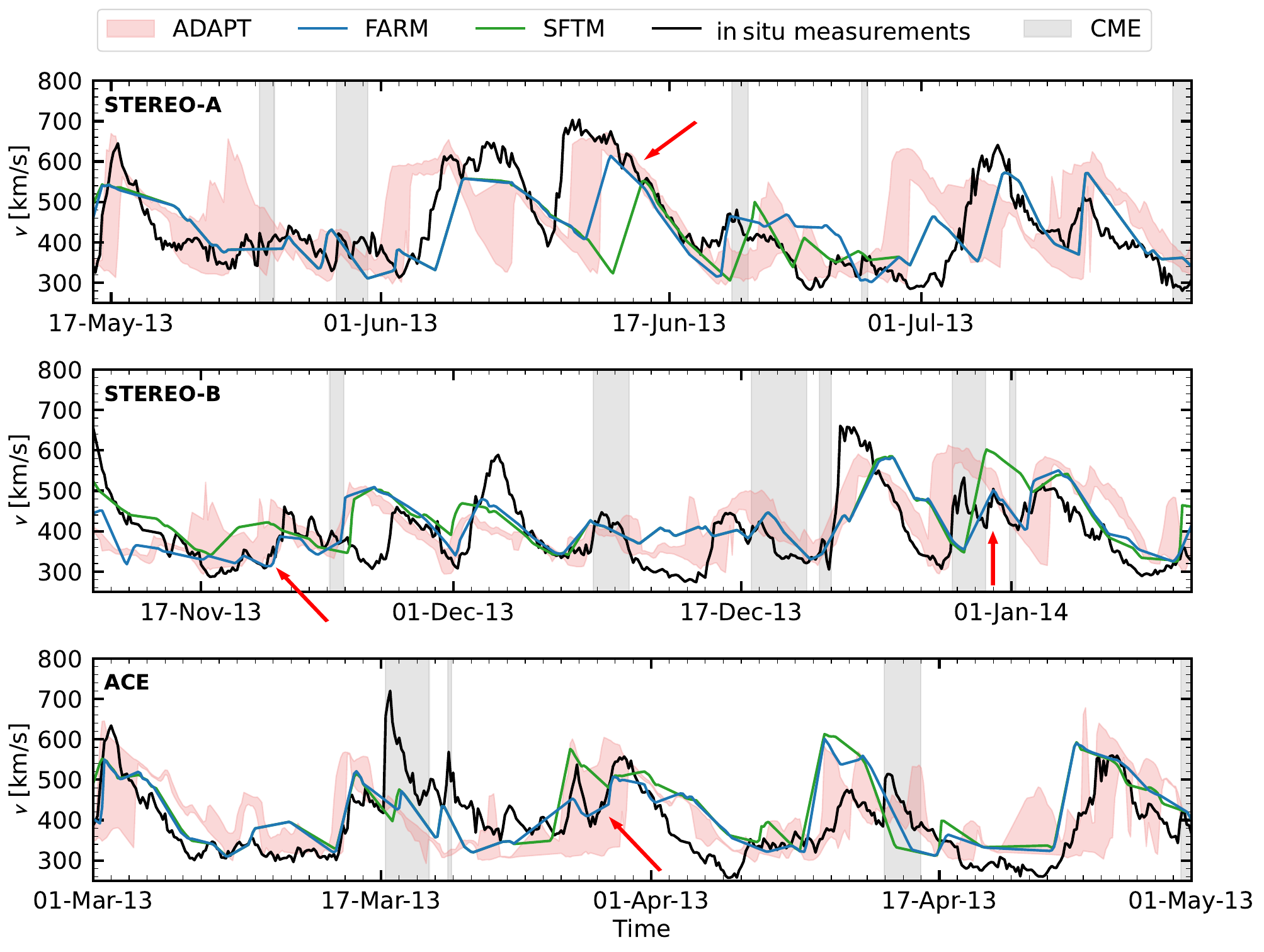}
   \caption{Modeled and observed in situ solar wind velocities are shown for the following time periods: 16 May to 16 July {2013} for STEREO-A, 11 November 2013 to 11 January 2014 for STEREO-B, and 1 March to 1 May 2013 for ACE. The observed in situ velocity, is shown in black. Model predictions based on FARM and SFTM maps are represented in blue and green, respectively. The red shaded region indicates the range of solutions from the ADAPT ensemble. Gray shaded areas denote time intervals identified as CMEs in the Richardson and Cane ICME catalog (ACE) or L. Jian’s ICME lists (STEREO-A, -B). Red arrows highlight periods where FARM results visibly outperform SFTM results.}
              \label{fig:insitu}%
\end{figure*}

\begin{table}[h]
\caption{{Solar wind velocity metrics for the time intervals shown in Figure~\ref{fig:insitu}.}}\label{tbl:stats_small}
\centering
\setlength{\tabcolsep}{6pt}
\renewcommand{\arraystretch}{1.2}
\small

\begin{tabular}{l | c c c | c c c}
\hline
& FARM & SFTM & ADAPT & FARM & SFTM & ADAPT \\
\hline
& \multicolumn{3}{c|}{$cc_{\mathrm{Pearson}}$} & \multicolumn{3}{c}{RMSE [km/s]} \\

\hline
STEREO-A & 0.63 & 0.50 & 0.31--0.54 & 83 & 95 & 97--118 \\
STEREO-B & 0.52 & 0.52 & 0.39--0.55 & 74 & 79 & 75--86 \\
ACE      & 0.60 & 0.56 & 0.33--0.57 & 75 & 83 & 78--96 \\
\hline

\hline
& \multicolumn{3}{c|}{MAE [km/s]} & \multicolumn{3}{c}{DTW} \\
\hline
STEREO-A & 63 & 67 & 71--82 & 5.9 & 8.3 & 7.6--10.4 \\
STEREO-B & 56 & 61 & 57--67 & 9.4 & 9.2 & 11.3--12.9 \\
ACE      & 59 & 67 & 61--78 & 8.1 & 9.4 & 9.1--12.3 \\
\hline
\end{tabular}

\end{table}

We run the \wsa\ model over a two-year period using each of the three types of magnetic maps as input. For this analysis, we selected a time frame between 2013 and {2014}, during which both {Solar TErrestrial RElations Observatories} \citep[STEREO:][]{2008kaiser_STEREO} were operational and positioned at separation angles ranging between $130^{\circ}$ and $170^{\circ}$ from Earth to obtain optimal coverage around the Sun. The \wsa\ model predicts solar wind bulk flow speed at the location of both STEREO spacecraft (A-head and B-ehind) as well as Earth. A more detailed description can be found in the Appendix~\ref{sec:wsa}. 

To evaluate the model results, we use solar wind in situ measurements from the {Advanced Composition Explorer} \citep[ACE:][]{1998stone_ACE}{}{} for Earth's location. For STEREO, we use plasma data from {PLAsma and SupraThermal Ion Composition} instrument \citep[PLASTIC:][]{2008galvin_PLASTIC}{}{}. {For consistency when making comparisons, the observed solar wind data (originally at 1-hour resolution) and the \wsa\ model output (provided on a non-uniform time grid) were both interpolated to a uniform 3-hour cadence.} We remove all CME intervals provided in the Richardson and Cane ICME (interplanetary CME) list\footnote{\url{https://izw1.caltech.edu/ACE/ASC/DATA/level3/icmetable2.htm}} \citep[RC-list:][]{2010richardson_RC-list}{}{} from ACE and the respective model data. For the two STEREO spacecraft, we use the ICME list from L. Jian\footnote{\url{https://stereo-ssc.nascom.nasa.gov/pub/ins_data/impact/level3/}}\citep[][]{Jian2018_iCME_catalog}{}{}. A two month segment of the in situ time series, including model results at each location, is shown in Figure~\ref{fig:insitu}. \\

{Using both the measured and predicted solar wind data, we conduct standard classification performance evaluation and time series correlation. We compute the root mean square error (RMSE), the mean absolute error (MAE), the {Pearson} correlation coefficient as well as a Dynamic Time Warping (DTW) analysis \citep{2022Samara}. DTW is a distance-based similarity measure used to compare two time series that may vary in speed or timing. It identifies an optimal nonlinear alignment between the time series by minimizing the cumulative distance between them. In this context, lower DTW values indicate better agreement between the model and observations, as they reflect smaller discrepancies in both the shape and timing of the profiles. More details are shown in  Appendix~\ref{app:stats}.} 

{Table~\ref{tbl:stats_small} presents the solar wind velocity performance metrics for the three time intervals shown in Figure~\ref{fig:insitu}. Across all spacecraft (STEREO-A, STEREO-B, and ACE) and nearly all metrics—Pearson correlation coefficient, RMSE, MAE, and DTW—the runs using FARM as input consistently outperform runs using both SFTM and ADAPT maps. Notably, FARM maps achieve higher correlation values and lower error metrics, indicating better agreement with in situ observations. We further quantify this performance over a longer, two-year period later in the manuscript.}\\

To qualitatively evaluate the FARM magnetic maps for advanced 3D heliospheric modeling, we use the 3D magnetohydrodynamic (MHD) heliospheric model \euhforia;  \citep[][]{pomoell2018}. It is a space weather modeling suit that utilizes a blend of empirical and physics-based methodologies to compute the temporal evolution of the inner heliospheric plasma environment. It comprises two primary components: a coronal model and a heliosphere model with the option to incorporate various configurations of coronal mass ejections. The coronal model generates data-driven solar wind plasma parameters at 0.1 AU through the construction of a magnetic field model representing the large-scale coronal magnetic field, along with the application of empirical relationships to ascertain plasma characteristics like solar wind speed and mass density. These parameters serve as boundary conditions to drive a 3D MHD model of the inner heliosphere, which extends from 21.5 $R_{sun}$ up to 2 AU. More details regarding the model setup can be found in the Appendix~\ref{sec:euhforia}.

\section{Results}

\subsection{Statistical Evaluation of \wsa Solar Wind Forecast}
\label{SubS:res_stats}

\begin{figure*}[h]
   \centering
   \includegraphics[width=1\linewidth]{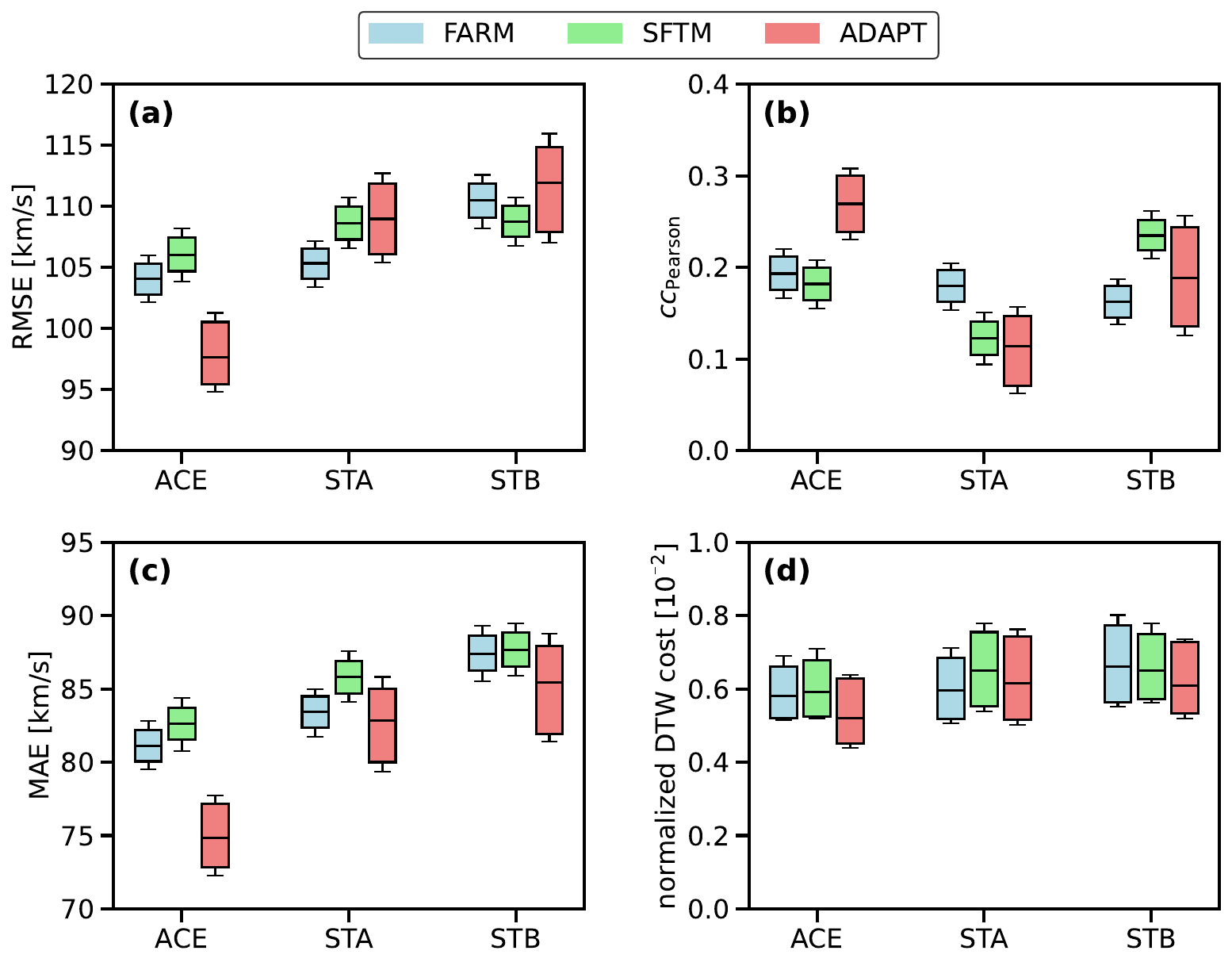}
   \caption{Statistical evaluation of two years of \wsa\ results from FARM, SFTM, and ADAPT maps for the location Earth (ACE), STEREO-A (STA) and STEREO-B (STB) respectively. Panel a and c show the RMSE and MAE between model results and in situ measurements, panel b depicts the {Pearson} correlation coefficient, and panel d the normalized DTW distance. The bars and error bars represent the $80\%$ and $95\%$ confidence intervals, respectively, while the horizontal bar indicates the median value.}
              \label{fig:stats}%
\end{figure*}

From two years of predicted solar wind data using the \wsa\ model with FARM, the SFTM, and ADAPT providing the global photospheric magnetic field maps, we derived the overall agreement between models and observations using the {Pearson} correlation coefficient ($cc_{\mathrm{Pearson}}$). We found that FARM performs between $6-50\%$ better than the base SFTM for STEREO-A and ACE over the full time period, but worse for STEREO-B. The reference maps from ADAPT, which are tuned to work well with \wsa\, best performs for ACE but is worse than FARM at STEREO-A. This is visualized in Figure~\ref{fig:stats}b.

To assess the agreement between the model results and in situ observations, we calculated the root mean square error (RMSE) and mean absolute error (MAE) of the solar wind speed, as defined in Equations~\ref{eq:MAE} and~\ref{eq:RMSE} (in Appendix~\ref{app:stats}). FARM and SFTM show very similar performance in terms of both RMSE and MAE, with FARM achieving values up to $3\%$ lower compared to SFTM. For the STEREO-A and -B spacecraft, FARM outperforms ADAPT in terms of RMSE by up to $3.5\%$, but shows a higher MAE. As expected, ADAPT performs best for ACE. These results are illustrated in Figure~\ref{fig:stats}a,c. Overall, all model results are within $15\%$ of each other, with ADAPT exhibiting the greatest variability due to its 12 different realizations.

Figure~\ref{fig:stats}d shows the normalized DTW distance, revealing that all three models produce similar results across all locations. FARM maps perform equal to or better than SFTM maps for all locations, with FARM performing up to $9\%$ better than SFTM and up to $5\%$ better than ADAPT for STEREO-A. ADAPT performs best for ACE. However, the confidence intervals are large and overlap, making it difficult to draw definitive conclusions about the relative performance of the models.
All statistical results are listed in Table~\ref{tbl:stats}.

\subsection{The 3D Solar Wind Structure}
      \label{S-3dcase}   

\begin{figure*}[h]
   \centering
   \includegraphics[width=1\linewidth]{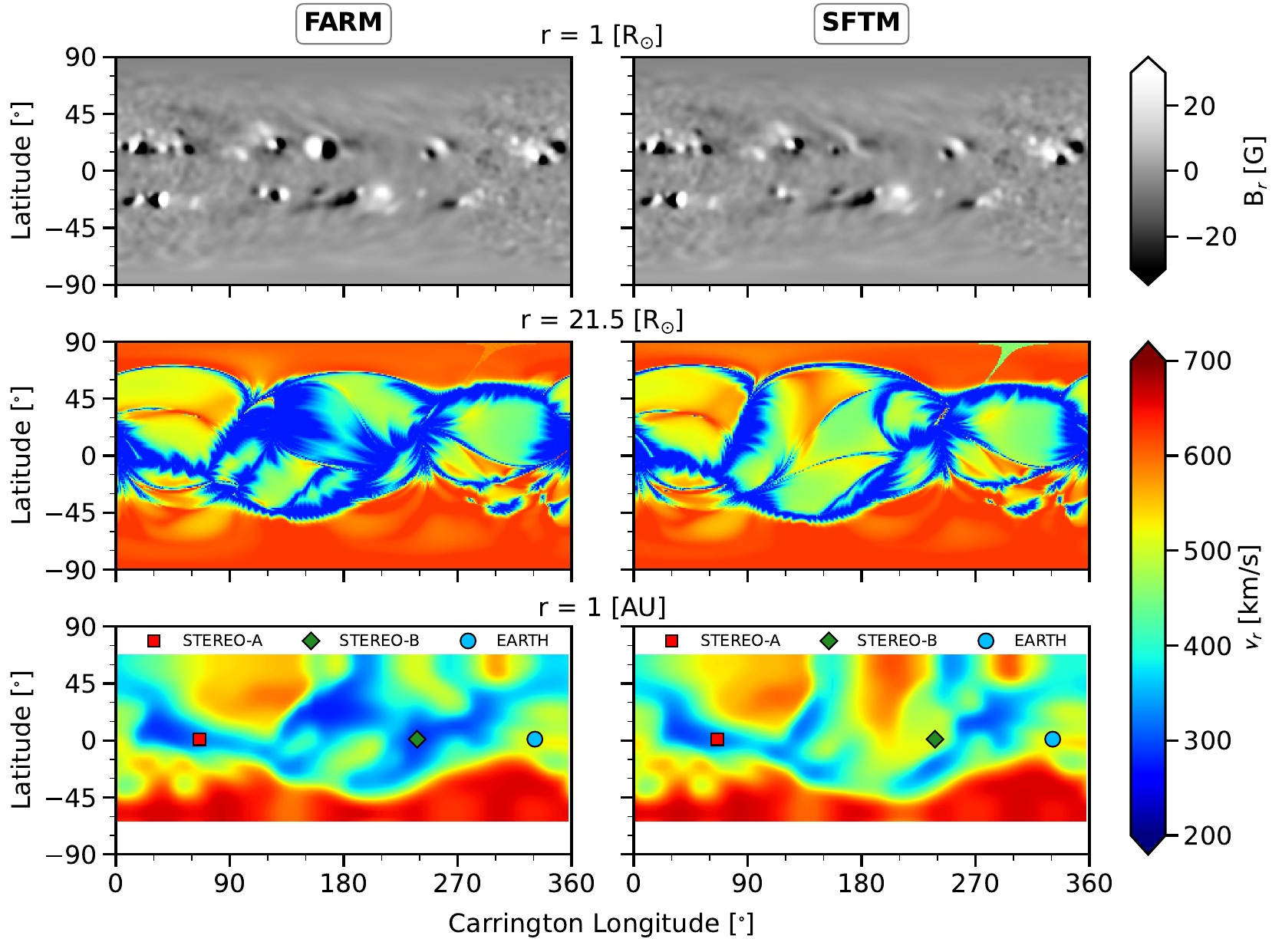}
   \caption{\euhforia\ model results using FARM (right column) and SFTM magnetic maps (left column) from 8 June 2011. From top to bottom the panels depict the photospheric magnetic field that was used as input to the coronal model, the solar wind at 0.1 AU which is the input to the heliospheric model and the solar wind results at 1 AU. {The simulation results correspond to 15 June 2011 23:59 UT} The blue, red, and green dots denote the projected locations of Earth, STEREO-A, and STEREO-B, respectively.}
              \label{fig:mhdruns}%
\end{figure*}
\begin{figure*}[h]
   \centering
   \includegraphics[width=1\linewidth]{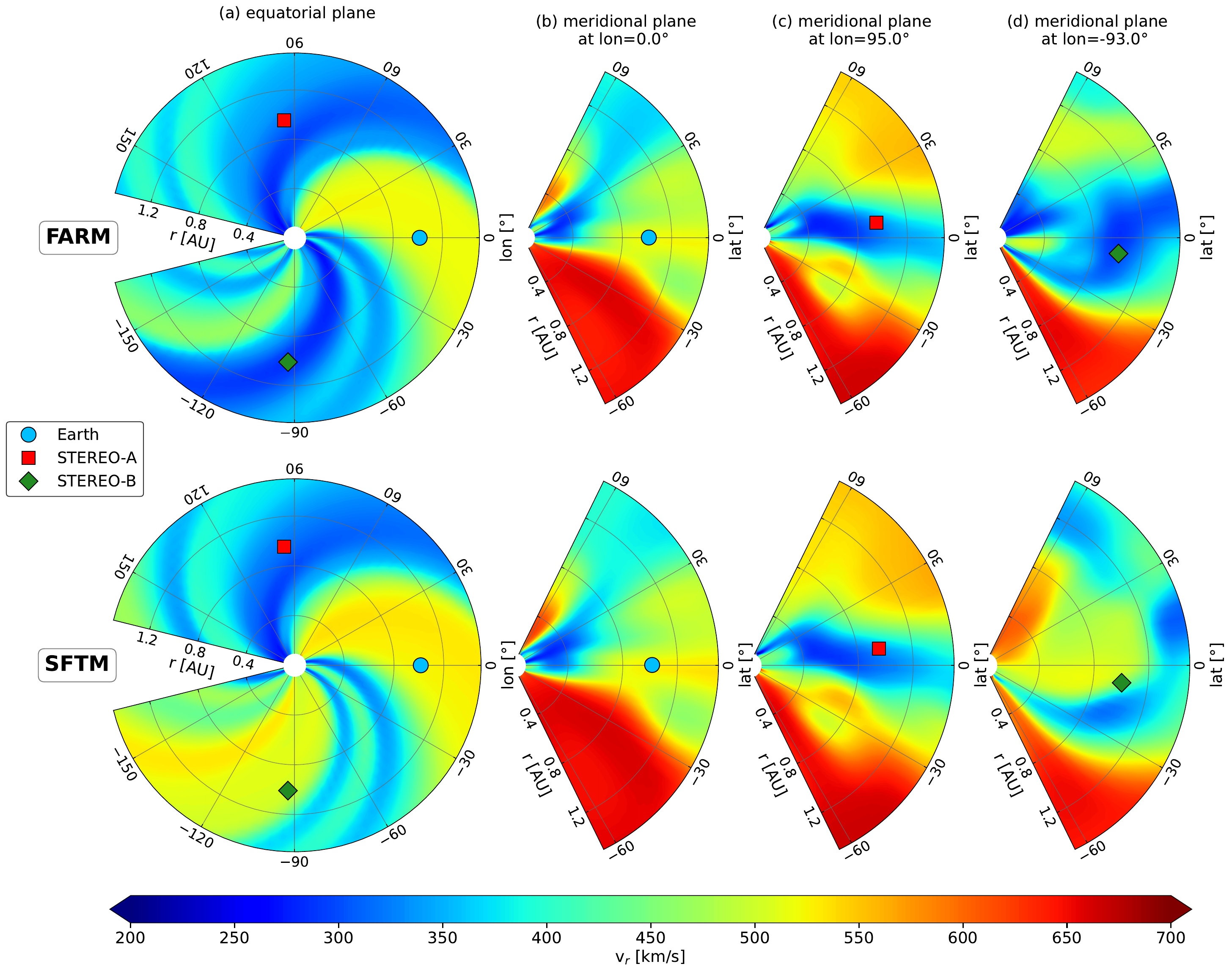}
   \caption{\euhforia\ model results of the same run as Figure~\ref{fig:mhdruns} but showing the equatorial plane (column a) and the meridional planes at the position of Earth (column b), at STEREO-A (column c) and STEREO-B (column d). {The simulation results correspond to 15 June 2011 23:59 UT.} FARM results are shown on the top row and SFTM results on the bottom row. The blue, red, and green dots denote the projected locations of Earth, STEREO-A, and STEREO-B, respectively. Note that the longitudes are given in Stonyhurst coordinates for clarity.}
              \label{fig:mhdcuts}%
\end{figure*}

\begin{figure*}[h]
   \centering
   \includegraphics[width=1\linewidth]{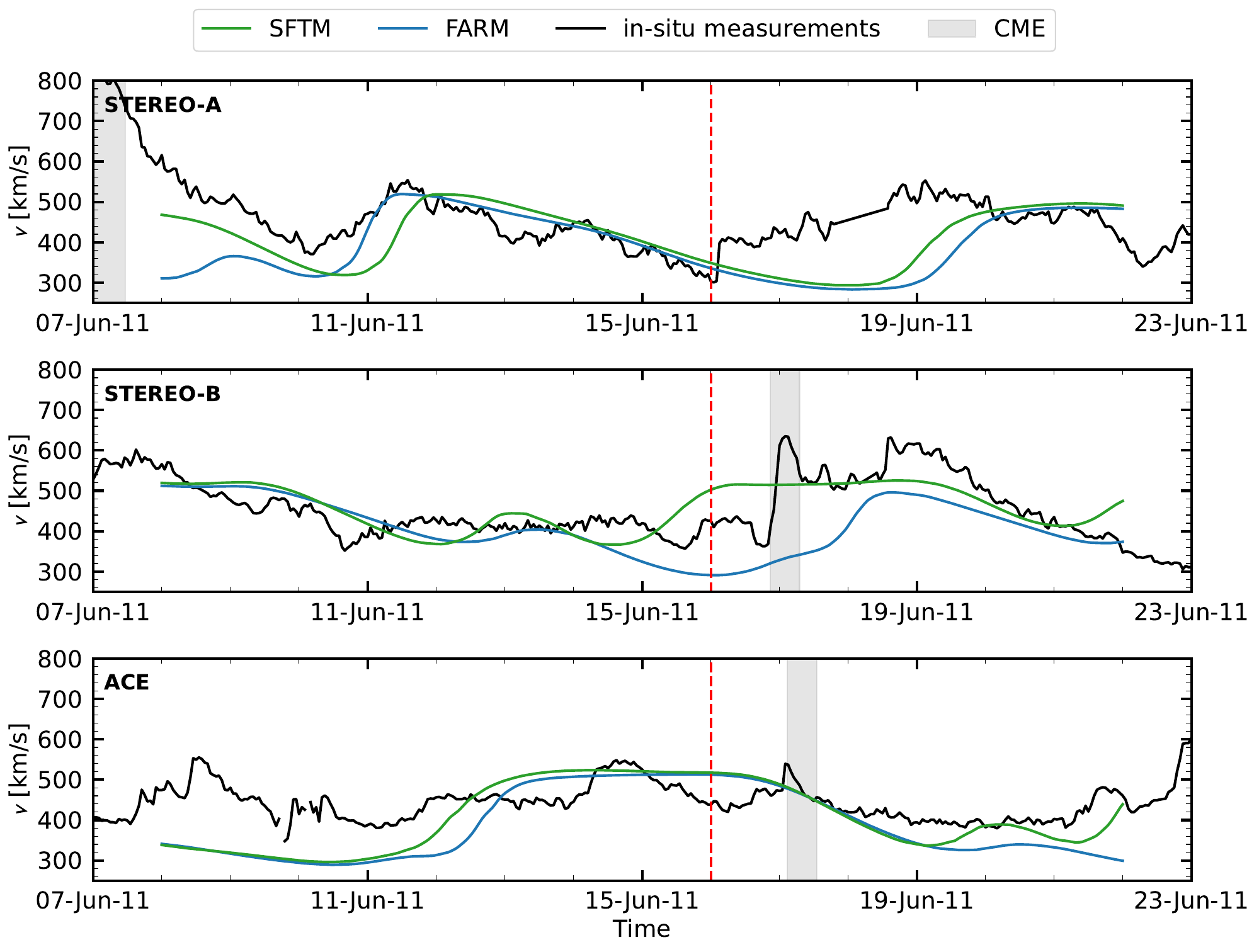}
   \caption{\euhforia\ solar wind predictions for  STEREO-A (top panel), STEREO-B (middle panel), and Earth (bottom panel) using FARM (blue line) and the SFTM (green line) results in comparison to the respective in situ measured data (black line). The gray shaded areas denote time intervals identified as CMEs in the Richardson and Cane ICME catalog (ACE) or L. Jian’s ICME lists (STEREO-A, -B). {The vertical red line marks 15 June 2011 23:59 UT, the date at which the simulation results are shown in Figures~\ref{fig:mhdruns} and~\ref{fig:mhdcuts}.}}
              \label{fig:in-situ_mhd}%
\end{figure*}

In Section~\ref{S-Statistics}, we conducted a statistical evaluation of magnetic maps incorporating far-side active regions using the \wsa\ solar wind model. While this method enables the mapping of solar wind speed at specific locations in the heliosphere (e.g. Earth, STEREO-A, and -B), it does not facilitate an assessment of the 3D heliospheric structure. To illustrate potential changes in the overall structure, changes that are not only relevant to mapping the background solar wind structure but are also capable of significantly affecting the propagation behavior of solar transients, we utilized the heliospheric MHD model \euhforia.

We ran \euhforia\ using a FARM and a SFTM magnetic map from 8 June 2011. This date was selected because it includes multiple far-side active regions, thus exhibiting significant differences in the photospheric far-side magnetic field. {The top panels of Figure~\ref{fig:mhdruns} show} the input maps, where FARM features large active regions that are absent from the SFTM map. {These discrepancies directly affect the coronal modeling results, as evidenced by differences in the solar wind speed at 0.1 AU (21.5 R$_\odot$).}The most notable difference appears in the SFTM results, which display a region of fast solar wind around $120$--$150^{\circ}$ Carrington longitude and approximately $45^{\circ}$ latitude, a feature not present in the FARM results. This suggests that the HSS observed in the SFTM case may be an artefact caused by incomplete magnetic field information, as this structure is absent in the FARM simulation and also not seen at 1 AU.

Due to these far-side differences, the solar wind profiles at 1 AU show significant variation in the Carrington longitudes between $30$--$150^{\circ}$, while appearing more similar near the Earth-facing (front-side assimilation) region, an expected outcome. Additionally, we find that the solar wind streams encountered by Earth, STEREO-A, and STEREO-B are generally consistent between the two models, although the 3D structure significantly differs in certain regions.

To provide a clearer view of the results, Figure~\ref{fig:mhdcuts} presents the equatorial plane along with the meridional planes at the locations of Earth, STEREO-A, and STEREO-B. As already seen in the radial planes (Figure~\ref{fig:mhdruns}), the most significant differences appear near the solar {far-side}, where FARM clearly shows the absence of a HSS in the northern hemisphere, visible in both the equatorial plane and the meridional plane at the location of STEREO-A. In contrast, the meridional planes at Earth and STEREO-B reveal largely similar modeled solar wind speeds, particularly at the spacecraft locations.

This is further confirmed in Figure~\ref{fig:in-situ_mhd}, which shows the \euhforia\ solar wind predictions as a function of time at Earth, STEREO-A, and STEREO-B. While the modeled solar wind from both magnetic maps aligns reasonably well with the in situ measurements, and with each other, for most of the time period, there are some differences, particularly in the STEREO-B results. {Here, SFTM predicts an earlier onset and a longer duration of a HSS compared to FARM results. However, due to the presence of a CME in the in situ data during that period, a definitive assessment of whether FARM or SFTM aligns better with observations is not possible—though the results from FARM appear more consistent with the expected solar wind structure.}


\section{Discussion}
      \label{S-disc}   

During periods of enhanced solar activity, the prediction of the solar wind speed is particularly difficult due to the complexity of the solar magnetic field configuration, enhanced by the frequent emergence of active regions out of Earth's field-of-view. Operative solar wind models generally perform at a long-term correlation of $cc = 0.2-0.5$ \citep{Reiss2020}, strongly varying between solar minimum and maximum. \cite{Milosic2023} evaluated the state-of-the-art operative empirical model Empirical Solar Wind Forecast (ESWF) 3.2 to have a correlation of $cc = 0.31$ during solar maximum. All our model runs were found to perform in the range of $cc \approx 0.11-0.27$. Although over short time periods, the correlation can easily exceed $cc = 0.5$. As expected, FARM and SFTM results are most similar at Earth, as the assimilated data is most accurate. Unexpectedly, FARM does not outperform SFTM at STEREO-B but rather on both STEREO-A and Earth. Used as a reference model, the ADAPT maps perform best for Earth. Here we note again that \wsa\ is tuned tuned for initiation with ADAPT maps and in situ observations at Earth's position, and therefore expected to perform better than other maps. {In addition to comparing model performance, it is important to note the general trend in observational data quality and model accuracy across spacecraft. Performance metrics such as MAE and RMSE are consistently best for ACE, followed by STEREO-A, and then STEREO-B. This ordering aligns with the availability of magnetic field information: ACE benefits from direct Earth-side observations, STA corresponds to regions that have just rotated out of Earth view and therefore has partial, more recent Earth-based magnetic field input, while STB observes far-side regions where the magnetic field information is oldest, except for updates incorporated through FARM. This gradient in data currency and quality naturally influences the model accuracy at each spacecraft location.}

\cite{MacNeice2009} and \cite{jian15} derived RMSE for solar wind predictions at Earth of $100-150$ km s$^{-1}$, which we are able to match in our study for all three locations, as well as input magnetic maps. ADAPT performs worst at STEREO-B at an RMSE of $112$ km s$^{-1}$ and best at $98$ km s$^{-1}$ at Earth. FARM and SFTM perform very similarly across all locations within $\pm5~\mathrm{km\,s^{-1}}$ of each other in the RMSE range of $100$–$110~\mathrm{km\,s^{-1}}$.
 This is also in agreement with \cite{Milosic2023} who found a RMSE of $110$ km s$^{-1}$.

\citet{2022Samara} established DTW as a reliable method for comparing time series in heliospheric physics. However, in our case, we find a large spread and very similar values across all models and locations, although there is a slight indication of improved performance from the FARM maps compared to SFTM.

\cite{2013Arge} were the first to implement far-side active regions in a flux transport model for solar wind forecasting. By inserting active regions into two Carrington rotations, they achieved a qualitative improvement of the forecast using \wsa\ runs. Their results agree with those derived in this study. {\citet{Knizhnik2024_farside_sw} analyzed the AFT-304 model, which incorporates far-side information via STEREO He {\sc ii} observations, and found that even when far-side active region emergence was clearly inferred, the differences in in situ solar wind predictions were minimal in most cases. Only under specific conditions did the inclusion of this data lead to noticeable differences in the predicted solar wind velocity. This selective impact is similar to our own findings, with overall improvements in ecliptic-plane forecasts remaining generally modest. Similarly, our results support their conclusion that a far-side active region must significantly displace open flux or alter magnetic connectivity in order to meaningfully affect solar wind conditions at observational points. \citet{Perri2024} further support this interpretation, demonstrating that even large far-side active regions may have negligible impact on solar wind conditions at Earth. Using a potential-field source-surface (PFSS) modeling of NOAA 12803 during Carrington rotaion (CR) 2240, they found minimal differences in solar wind speed and coronal structure when comparing simulations with and without a inserted far-side bipole. } 

While the improvement in predicted solar wind speeds within the ecliptic plane is relatively small, substantial differences become evident when considering the 3D structure. Large-scale changes in solar wind structure (such as absent or present HSSs) are possible, as illustrated in Figures~\ref{fig:mhdruns} and~\ref{fig:mhdcuts}. These changes may not directly impact the solar wind within the ecliptic, but they can significantly influence the propagation of CMEs in heliospheric models. CMEs can be notably deformed or deflected by the presence or absence of HSSs, meaning that even variations in solar wind structure outside the ecliptic can substantially affect predictions of geomagnetic storms at Earth or other planets. {One mechanism contributing to these off-ecliptic structural differences is the dynamic evolution of coronal holes and active regions. \citet{Wang2024_farside} highlighted that the influence of newly emerged ARs is strongly modulated by their location, polarity, and interaction with preexisting coronal holes. While the total dipole or quadrupole moment may remain stable, interchange reconnection can still locally reshape coronal hole boundaries, causing latitudinal shifts in HSS footprints or temporary modifications to wind speed. This helps explain the cases in our study where FARM and SFTM magnetic maps produce qualitatively different solar wind structures off the ecliptic, even if in-ecliptic differences remain modest.}

Generally, we do not expect significant improvements in background solar wind prediction from the inclusion of far-side active regions in magnetic maps over extended periods. Instead, the impact is likely to be more pronounced during shorter, {more localized,} intervals when the solar {far-side} evolves differently from the front side, i.e., when the magnetic field structure on the {far-side} deviates from that assumed by flux transport models or traditional synoptic charts.

Another reason for the relatively modest improvements in ecliptic solar wind predictions is the typical location of active regions away from the solar equator (up to $\approx 35^{\circ}$ latitude depending on the solar cycle; \citealt{2023Weber}). Larger improvements might be expected near the end of solar maximum, when active regions tend to emerge at lower latitudes, but, during solar minimum, little to no difference between FARM and SFTM results is expected.

It is also worth noting that the inserted far-side active regions are assumed to have zero net magnetic flux. This means that the overall flux balance of the magnetic map remains unchanged, and therefore, minimal large-scale restructuring may occur. Finally, active regions generally do not emerge within large open-field regions, i.e., coronal holes which are the source regions of HSSs \citep{cranmer2002,2018heinemann_paperI,2020Heinemann_chevo}. As a result, cases in which the presence or absence of far-side active regions leads to significant changes in HSSs are likely the exception rather than the rule.

\section{Summary and Outlook}
      \label{S-sum}   
In this study, we evaluated magnetic maps, which include far-side active regions derived from helioseismic measurements, as input to the \wsa\ solar wind prediction model, as well as to the heliospheric MHD model \euhforia. \\

Our findings can be summarized:
\begin{itemize}
    \item Both FARM and SFTM align well with state-of-the-art magnetic maps like ADAPT, delivering reliable solar wind forecasts in the ecliptic plane during the tested solar maximum period and serving as viable inputs for the \wsa\ and \euhforia\ models.
    
    \item Using the \wsa\ model, we demonstrate that magnetic maps that incorporate far-side information improve long-term solar wind predictions for 2013–2014 by up to $6$--$50\%$ in terms of the {Pearson} correlation coefficient, and by up to $3\%$ for RMSE and MAE, particularly at the locations of Earth and STEREO-A.

    \item {Using advanced 3D modeling, we highlight how changes in the far-side photospheric magnetic field can significantly alter the solar wind structure, potentially enhancing the geoeffectiveness of solar transients.}

    \item {While localized and global differences in solar wind structure can be substantial, particularly off the ecliptic, improvements in predicting solar wind speed at Earth remain moderate, reflecting the limitations of current models and the complex influence of far-side magnetic activity.}

\end{itemize}

The evaluation in this study focused solely on the improvement of FARM from the base SFTM due to the inclusion of the far-side active regions, as described in \cite{Yang2024_FARM}, without deviation from the default model settings. In this article, we show the benefits of the inclusion of far-side information over a standard surface flux transport model. Going forward we will take further steps, for example, optimizing surface flux transport model parameters such as surface flows or diffusion, adjusting the settings of the solar wind model, and even using the methodology of detecting and inserting the far-side active regions detected using helioseismic holography \citep[see][]{Yang2023} in collaboration with more advanced surface flux transport models. We expect that the resultant new global solar magnetic maps will contribute significantly to advancing heliospheric research and improving space weather forecasting.

%

\appendix   

\section{Model Setup}
\label{app:models}

\subsection{Wang-Sheeley-Arge Model \wsa}
\label{sec:wsa}
The \wsa\ model is composed of three primary stages. First, the PFSS model calculates the inner coronal magnetic field by extrapolating from the photosphere to the source surface (R${\mathrm{SS}}$). This is achieved via a spherical harmonic expansion that matches the observed photospheric flux and imposes a radial magnetic field at R${\mathrm{SS}}$. Next, the solution is extended outward from the Schatten current sheet (SCS) radius (R${\mathrm{SCS}}$) to 5 solar radii using the SCS model. This method simulates the formation of a current sheet and helmet streamer above the magnetic neutral line by temporarily assuming a unipolar radial field and resolving a potential field to preserve open field lines and minimize the energy of the system.. Finally, the magnetic field at 5 solar radii is used to estimate the  solar wind speed through an empirical relationship. {This empirical relationship used to estimate solar wind speed commonly depends on two key parameters: the magnetic flux tube expansion factor ($f_{\mbox{\scriptsize exp}}$; e.g.  \citealt{wang90,1992wang,1996wang}) and the angular distance from the coronal hole boundary ($\Theta_B$ or DCHB; e.g.  \citealt{2001Riley,Rileyetal2015}). The expansion factor $f_{\mbox{\scriptsize exp}}$ is inversely correlated with solar wind speed and is associated with the modulation of outflow velocities along open magnetic field lines, particularly in wave–turbulence-driven acceleration models. In contrast, $\Theta_B$ measures the angular separation between a field line footpoints and the nearest coronal hole boundary, serving as a proxy for the likelihood of magnetic reconnection processes.}

{Recent studies \citep{2012riley_CORHEL,Rileyetal2015,Wallace2020} indicate that the traditional inverse relationship between $f_{\mbox{\scriptsize exp}}$ and solar wind speed breaks down in certain magnetic topologies, such as pseudostreamers. Consequently, modern implementations of the WSAmodel, including the version used in this
study, place greater emphasis on the coronal hole boundary distance parameter ($\Theta_B$).}

In this study we used R$_{\mathrm{SS}} =2.51$R$_{\odot}$, R$_{\mathrm{ScS}} =2.49$R$_{\odot}$ with the empirical solar wind relation:
\begin{equation}
    V = V_0 + \frac{V_m}{(1+f_{\mbox{\scriptsize exp}})^{C_1}}\,\left(1 - C_2\,\mbox{exp}\left[-\left(\frac{\Theta_B}{C_3}\right)^{C_4}\right]\right)^{C_5}
    \label{eqn:wsa_windspeed}
\end{equation}
where $V_0$, $V_m$, and $C_1$ - $C_5$ are parameters that were empirically derived. They are listed in Table~\ref{table:wsa_parameters}.

\begin{table}
\qquad \qquad \,
  \begin{tabular}{|c|c|c|c|c|c|c|}
    \hline
    $V_0$ & $V_m$ & $C_1$ & $C_2$ & $C_3$ & $C_4$ & $C_5$ \\
    \hline 
    286 m/s& 625 m/s& 2/9 & 0.8 & $1$ & 2 & 3 \\
    \hline
  \end{tabular}
  \caption{Parameter values used for solar wind speed predictions in Equation \ref{eqn:wsa_windspeed}.}
  \label{table:wsa_parameters}
\end{table}

\subsection{EUropean Heliospheric FOrecasting Information Asset: \euhforia} \label{sec:euhforia}

\euhforia\ is a computational framework designed to simulate large-scale solar wind and magnetic field behavior within the inner heliosphere. Starting from the inner boundary at a heliocentric distance of $r_H = 0.1$ au, the evolution of plasma and magnetic structures for $r > r_H$ is governed by numerically solving the 3D magnetohydrodynamic (MHD) equations, as described in \citet{pomoell2018}.\\

The solar wind background at $r = r_H$ can be initialized using several methods. In this study, a semi-empirical approach, similar to that of the \wsa\ model (refer to \ref{sec:wsa}), is used. Specifically, the coronal magnetic field from the solar surface up to $r_H$ is reconstructed using a two-part system: a PFSS model for the lower corona and a Schatten current sheet (SCS) model for the upper corona. These are implemented using the method of expansion in solid harmonics up to degree l=140 to solve the Laplace equation, with the PFSS model extending to a source surface radius of $2.6$ R$_\odot$ and the current sheet model beginning at a lower height of $2.3$ R$_\odot$. This parameter configuration is a similar approach to that of the \wsa\ model and is known to significantly influence the structure of the extrapolated magnetic field \citep[e.g.][]{McGregor2008, 2019asvestari}. The numerical grid used includes a radial resolution of approximately $\Delta r \approx 0.02$ R$_\odot$ for the PFSS and $\Delta r \approx 0.15$ R$_\odot$ for the SCS domain, with an angular resolution of $1^\circ$, matching the input magnetograms.

Magnetic field lines are traced through the coronal model to assess connectivity, and a map of the angular distance $d$ to the closest open field line is generated. The solar wind speed is then assigned using the empirical relation:
\begin{equation}
    V = V_\mathrm{s} + V_\mathrm{f}\frac{1}{(1+f)^{\alpha}} \left( 1 - 0.8e^{(-\frac{d}{w})^{\beta}} \right)^{3}
\end{equation}

where $V_\mathrm{f} = 675$ km/s, $V_\mathrm{s} = 240$ km/s, $\alpha = 0.22$, $\beta = 1.25$, and $w =$ \SI{0.01}{\radian} and $f$ is the flux tube expansion factor, similar to the form used
by \citet{2011mcgregor}.


The MHD simulation is executed over a 10-day period to achieve a quasi-steady-state solar wind configuration (no CMEs are included). The computational domain spans a radial range from $0.1$ to $2.0$ au and latitudes from $-70^\circ$ to $70^\circ$, using a grid of $(512, 35, 90)$ cells in radial, latitudinal, and longitudinal directions, respectively.

\section{Statistics} \label{app:stats}

We calculate the mean absolute error (MAE) and the root mean square error (RMSE) between the modeled and the in situ measured solar wind velocities for the entire 2-year timeseries.
\begin{equation} \label{eq:MAE}
\mathrm{MAE} = \frac{1}{n} \sum_{i=1}^{n} |y_i - \hat{y}_i|
\end{equation}

\begin{equation} \label{eq:RMSE}
\mathrm{RMSE} = \sqrt{\frac{1}{n} \sum_{i=1}^{n} (y_i - \hat{y}_i)^2}
\end{equation}

where $n$ represents the number of data points (in the model) or observations, $y_i$ is the actual (observed) value for the $i$-th data point and $\hat{y}_i$ represents the predicted (modeled) value for the $i$-th data point.\\

{To quantify the similarity between time series from model output and in situ observations, we compute the {normalized Dynamic Time Warping (DTW) distance} using the \texttt{dtaidistance} Python package. DTW is a time-series comparison algorithm that aligns two sequences by minimizing the cumulative distance between them, allowing for temporal shifts and local stretching or compression. This makes DTW particularly well-suited for comparing solar wind profiles, which may exhibit similar shapes but be offset in time due to variations in propagation speed or arrival time.}

{Before computing DTW, each time series is standardized by subtracting its mean and scaling to unit variance. This ensures that the comparison emphasizes the shape of the time series rather than their absolute amplitude differences. The DTW distance is then computed between the standardized model and observational series.}

{To account for differences in time series length, we normalize the DTW distance by the number of data points, yielding a scale-independent similarity metric. In this formulation, a {lower normalized DTW value indicates greater similarity} between the time series in both pattern and timing, while a higher value reflects larger misalignments or structural differences.}

{To assess the robustness of the DTW results, we compute confidence intervals by varying the DTW window size, which controls the maximum temporal displacement allowed during alignment. This provides an estimate of the uncertainty associated with the DTW-based similarity metric.}

{This approach allows us to capture not only amplitude and shape differences, but also timing variations between modeled and observed solar wind features, making DTW a more flexible and informative metric than traditional point-by-point comparisons.}

 \begin{table}[h]
 \caption{{Solar wind velocity metrics} for the full time interval of two years, 2013 and 2014. The values in the brackets show the $95\%$ confidence intervals.}\label{tbl:stats}
 \begin{tabular}{l | c c c}     
 \hline
  & FARM & SFTM & ADAPT \\
  \hline
   &\multicolumn{3}{c}{$cc_{\mathrm{Pearson}}$}\\
  \hline
STEREO-A & 0.18 [0.16;0.20]&0.12 [0.10;0.14]&0.11 [0.07;0.15] \\
STEREO-B & 0.16 [0.15;0.18]&0.24 [0.22;0.25]&0.19 [0.14;0.24] \\
ACE & 0.19 [0.18;0.21]&0.18 [0.16;0.20]&0.27 [0.24;0.30] \\
\hline
&\multicolumn{3}{c}{RMSE [km/s]}\\
\hline
STEREO-A & 105.40 [104.18;106.59]&108.69 [107.36;110.11]&108.96 [106.11;111.82] \\
STEREO-B & 110.41 [109.03;111.76]&108.78 [107.40;110.13]&111.97 [108.09;114.87] \\
ACE & 104.03 [102.73;105.29]&105.93 [104.50;107.31]&97.69 [95.35;100.60] \\
\hline
&\multicolumn{3}{c}{MAE [km/s]}\\
\hline
STEREO-A & 83.44 [82.32;84.59]&85.86 [84.75;87.01]&82.82 [79.99;85.06] \\
STEREO-B & 87.30 [86.09;88.47]&87.68 [86.51;88.92]&85.44 [82.13;88.04] \\
ACE & 81.13 [80.05;82.14]&82.55 [81.42;83.75]&74.87 [72.81;77.23] \\
\hline
&\multicolumn{3}{c}{normalized DTW cost [$10^{-2}$]}\\
\hline
STEREO-A & 0.60 [0.51;0.71]&0.65 [0.54;0.78]&0.62 [0.50;0.76] \\
STEREO-B & 0.66 [0.55;0.80]&0.65 [0.56;0.78]&0.61 [0.52;0.74] \\
ACE & 0.58 [0.51;0.69]&0.59 [0.52;0.71]&0.52 [0.44;0.64] \\
 \hline
 \end{tabular}
 \end{table}

\begin{acknowledgments}
The SDO, ACE and STEREO data are available by courtesy of NASA and the respective science teams. We thank the International Space Science Institute (ISSI, Bern) for the generous support of the ISSI team “Quantitative Comparisons of Solar Surface Flux Transport Models” (2024–2025). The views expressed are those of the authors and do not reflect the official guidance or position of the United States Government, the Department of Defense or of the United States Air Force. 
\end{acknowledgments}

\begin{authorcontribution}
SGH led the manuscript preparation and conducted the data analysis. DY implemented and executed the FARM model. SIJ ran the \wsa\ model, while JP and EA carried out the \euhforia\ simulations. CJH provided the ADAPT maps. All authors contributed to the discussion of results and the final version of the manuscript.
\end{authorcontribution}

\begin{fundinginformation}
 DY acknowledges support from ERC Synergy grant WholeSun 810218.  This research was funded in whole, or in part, by the Austrian Science Fund (FWF) Erwin-Schr\"odinger fellowship J-4560. SGH acknowledges funding from the Research Council of Finland (Academy Fellowship): 370747 (RIB-Wind). JP and SGH acknowledge funding from the Academy of Finland project SWATCH (343581). CJH is partially supported by Air Force Office of Scientific Research (AFOSR) tasks 22RVCOR012 and 25RVCOR001. EA acknowledges support from the Finnish Research Council (Research Fellow grant number 355659).
\end{fundinginformation}

\begin{dataavailability}
Data generated during the current study are available from the corresponding author on reasonable request.
\end{dataavailability}

%
%
%

\end{document}